# Controllably releasing long-lived quantum memory for photonic polarization qubit into multiple separate photonic channels

**Lirong Chen, Zhongxiao Xu, Weiqing Zeng, Yafei Wen, Shujing Li and Hai Wang**

The State Key Laboratory of Quantum Optics and Quantum Optics Devices, Collaborative Innovation Center of Extreme Optics, Institute of Opto-Electronics, Shanxi University, Taiyuan, 030006, People's Republic of China

E-mail: wanghai@sxu.edu.cn

**Abstract**
We report an experiment in which long-lived quantum memories for photonic polarization qubits (PPQs) are controllably released into any one of multiple separate channels. The PPQs are implemented with an arbitrarily-polarized coherent signal light pulses at the single-photon level and are stored in cold atoms by means of electromagnetic-induced-transparency scheme. Reading laser pulses propagating along the direction at a small angle relative to quantum axis are applied to release the stored PPQs into an output channel. By changing the propagating directions of the read laser beam, we controllably release the retrieved PPQs into 7 different photonic output channels, respectively. At one of the output channels, the measured maximum quantum-process fidelity for the PPQs is 94.2% at storage time of $t=$ 0.85ms. At storage time of 6 ms, the quantum-process fidelity is still beyond 78%, the threshold for the violation of the Bell inequality. The demonstrated controllable release of the stored PPQs may extend the capabilities of the quantum information storage technique.

# 1. Introduction

Quantum networks (QNs), comprising of many quantum nodes and quantum channels, provide a crucial platform to perform scalable quantum information processing [1, 2]. Quantum nodes are used for processing and storing quantum information (qubits), while quantum channels are used for transporting quantum information between different nodes [1]. Flying photon qubits are good carriers of quantum information since they travel fast and weakly interact with environment [1, 2]. Cold atomic ensembles are promising matter nodes [1, 2] since long-lived and/or efficient quantum memories for single photon or photon qubits can be achieved via spontaneous Raman scattering [3-6], dynamic electromagnetic-induced-transparency (EIT) [7, 8]. Besides requiring the capacity to store quantum information in QN nodes with long lifetime and high efficiency, routing the retrieved photon qubits from an atomic memory into one of many output channels is also needed. For example, in some quantum information protocols such as quantum repeater with multiplexed memories [9, 10] and scalable quantum computing with atomic ensembles [11], the stored photons in atomic ensembles are required to be released into a desired quantum channel according to previous measurement results. By introducing optical switches into QNs, one can realize the retrieved-photon routing with multiple channels. However, the introduction of these optical switches will result in additional optical losses and also disturb the quantum states of the single photons. To avoid these questions, one can integrate the function of optical switches into quantum memory. In the past decades, quantum node with the ability to route single photons into a desired channel have been theoretically proposed and experimentally demonstrated in various physical systems such as cavity-QED system [12], circuit QED system [13], opto-mechanical system [14], waveguide-emitter system [15-18], and pure linear optical system [19] and so on [20]. For EIT-based light storage or slow-light systems, two- or three-channel optical routers of signal fields have been realized by applying control light beams propagating along different directions [21, 22] or switching on reading light beams operating on different wavelengths [23]. Recently, in a gradient-echo memory system, spatially addressable readout and erasure of an image has been experimentally demonstrated [24]. However, since the input light signals in these experiments [20-24] are intensive pulses, the quantum feature of these optical routers have not been characterized.

Here, we demonstrate an experiment in which the readouts of quantum memories for photonic polarization qubits (PPQs) can be routed into any one of multiple separate photonic channels. The PPQs are implemented with an arbitrarily-polarized coherent light (input signal) pulses at the single-photon level and co-propagate with a writing light beam through the cold atomic ensemble along z-axis. By means of EIT-based optical storage scheme [25-27], i.e., switching off a writing beam, we store the input PPQs as spin waves (SWs) in the cold atoms. For obtaining long-lived memories for PPQs, we apply a moderate magnetic field along z-axis to lift the Zeeman degeneracy [7] and then remove the fast decoherence coming from magnetic-field-sensitive SWs out of the memories. Reading-beam light pulses propagating along a direction at a small angle respect to the quantum axis defined by the magnetic field are applied to release the stored PPQs into an output channel. By changing the read-beam propagating directions, we release the stored PPQ into 7 different output channels, respectively. At one of the output channels the measured maximum quantum-process fidelity for the retrieved single-photon polarization states is 94.2% for storage time of $t=$ 0.85ms. At storage time of 6 ms, the quantum-process fidelity is ~80%, which is still beyond the threshold for the violation of the Bell inequality.

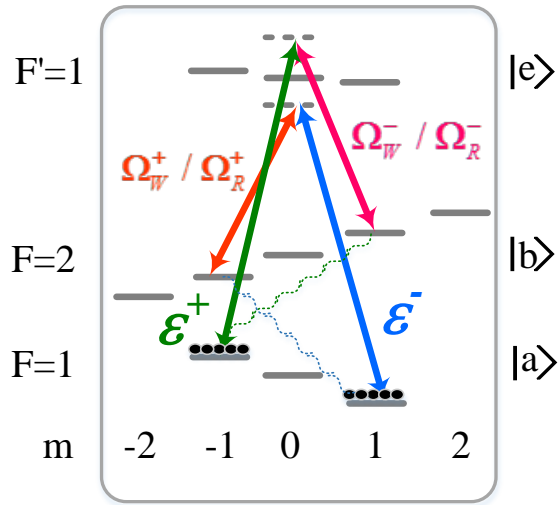

FIG. 1 (color online) Three-level Λ-type EIT system for the storage of the signal light fields in a moderate magnetic field ( $B_0 = 12.5G$ ). $\varepsilon^+$ and $\varepsilon^-$ denote right-circularly and left-circularly polarized components of the signal light field, respectively. $\Omega_W^+ / \Omega_R^+$ and $\Omega_W^- / \Omega_R^-$ denote right-circularly and left-circularly components of the linearly polarized writing/ reading light fields, respectively.

## 2. Theoretical model

The involved levels of $^{87}$Rb atoms is shown in Fig.1, where $|a\rangle = |5^2S_{1/2}, F=1\rangle$, $|b\rangle = |5^2S_{1/2}, F=2\rangle$ and $|e\rangle = |5^2P_{1/2}, F'=1\rangle$. The input signal and writing/reading coupling light fields are tuned to the transitions $|a\rangle \leftrightarrow |e\rangle$ and $|b\rangle \leftrightarrow |e\rangle$, respectively, their frequency difference is $\Delta\omega = \omega_{be} - \omega_{ae}$, which matches the resonance of the two-photon transition $|a\rangle \leftrightarrow |b\rangle$. The PPQs are implemented with an arbitrarily-polarized coherent light (input signal) field $\hat{\varepsilon}_{in}(t)$ at single-photon level. The signal field $\hat{\varepsilon}_{in}(t)$ is expressed by:

$$\hat{\varepsilon}_{in}(t) = c_1 |R\rangle + c_2 |L\rangle , \qquad (1)$$

where $|R\rangle$ and $|L\rangle$ denote right-circularly ($\sigma^+$) and left-circularly ($\sigma^-$) polarized components, respectively, $c_1$ and $c_2$ are their amplitudes with $|c_1|^2 + |c_2|^2 = 1$. The writing (reading) light field is vertical polarization and then can be viewed as the superposition of $\sigma^+$- and $\sigma^-$- polarized components $\Omega_W^+$ ($\Omega_R^+$) and $\Omega_W^-$ ($\Omega_R^-$). The quantum axis is defined by applying a bias magnetic field $B_0$ along z-direction. The atoms are initially prepared into Zeeman sublevels $|a_{m_a=-1}\rangle$ or $|a_{m_a=1}\rangle$ with equal population (*m* represents the magnetic quantum number) by optical pumping. By means of dynamic EIT process, i.e., switching off the writing beam, we can store the $\sigma^+-$ and $\sigma^--$ polarized components of the PPQ as two distinct SWs [7]. Under the condition of a weak bias magnetic field, both SWs include magnetic-field-sensitive and magnetic-field-insensitive SW components and thus storage lifetime is very short due to the fast decay of the magnetic-field-sensitive SWs [7]. In order to eliminate the bad influences of the magnetic-field-sensitive SWs, we follow the storage scheme demonstrated in Ref. [7], i.e., applying a moderate magnetic field $B_0$ on a cold-atom cloud to lift Zeeman degeneracy [as show in Fig. 1]. In this case, the magnetic-field-sensitive SW components are removed out of EIT storage systems and the PPQ can be only mapped on two magnetic-field-insensitive SWs $|\psi^\pm\rangle$, respectively [7]. The magnetic-field-insensitive SWs $|\psi^\pm\rangle$ are associated with the coherences $|m_a=\pm 1\rangle \leftrightarrow |m_b=\mp 1\rangle$, which can be expressed as:

$$|\psi^+\rangle = \sqrt{\frac{2F_a+1}{N}} \sum_{j=1}^{N_m} |a_1^{m_a=-1}\rangle...|b_j^{m_b=1}\rangle...|a_{N_{m_a}}^{m_a=-1}\rangle e^{-\vec{k}_A \cdot \vec{r}_j} \quad \text{and}$$

$$|\psi^-\rangle = \sqrt{\frac{2F_a+1}{N}} \sum_{j=1}^{N_m} |a_1^{m_a=1}\rangle...|b_j^{m_b=-1}\rangle...|a_{N_{m_a}}^{m_a=1}\rangle e^{-\vec{k}_A \cdot \vec{r}_j}, \quad (2)$$

respectively, where $\vec{k}_A = \vec{k}_S - \vec{k}_W$ is wave-vector of the two SWs $|\psi^\pm\rangle$, $\vec{k}_s$ and $\vec{k}_w$ are the wave-vectors of the signal and write fields, respectively. For suppressing the dephasing effect due to atomic motion [3,7], we let the input signal and writing beams collinearly go through the cold-atom cloud along z-axis, thus, $\vec{k}_s = \frac{\omega_{ae}}{c}\vec{e}_z, \vec{k}_w = \frac{\omega_{be}}{c}\vec{e}_z$, and the wavelength of SWs reach their maximal value $\lambda_a = c/\vec{k}_A$, where $\vec{e}_z$ is the unit vector along z-axis.

After a storage time *t*, we apply a reading light pulse to convert the two stored SWs into flying PPQ. As shown in Fig.2, a reading light field $\Omega_{Ri}$ propagates along the direction at a small angle of $\theta$ relative to z-axis, its wave-vector is written as: $\vec{k}_R(\theta) = \frac{\omega_{be}}{c}\cos\theta\vec{e}_z + \frac{\omega_{be}}{c}\sin\theta\vec{e}_x$, $\vec{e}_x$ is the unit vector along x-axis. The wave-vector of the retrieved signal photons $\vec{k}_{RS}(\theta')$ can be calculated according to the phase matching condition $\vec{k}_w - \vec{k}_s = \vec{k}_{R(\theta)} - \vec{k}_{RS(\theta')}$ [28], which is:

$$\vec{k}_{RS(\theta')} = \frac{\omega_{ae}}{c}\cos\theta'\vec{e}_z + \frac{\omega_{ae}}{c}\sin\theta'\vec{e}_x, \qquad (3)$$

where the angle $\theta' = arctg\frac{\sin\theta}{(\omega_{ae}-\omega_{be})/\omega_{be} + \cos\theta}$. For $^{87}$Rb atomic system, we can calculate $(\omega_{ae}-\omega_{be})/\omega_{be} \approx 10^{-5}$. So, for the case of $\theta \ll 5°$, we have $\cos\theta \gg 10^{-5}$, thus $\theta' \approx \theta$, which means that the retrieved signal photons approximately propagate along the same direction as that of the reading beam and then we can effectively collect the retrieved signal photons along the reading-beam direction. Based on this fact, we can release the stored PPQ into a desired output port by selecting an appropriate reading-beam propagating direction around z-axis.

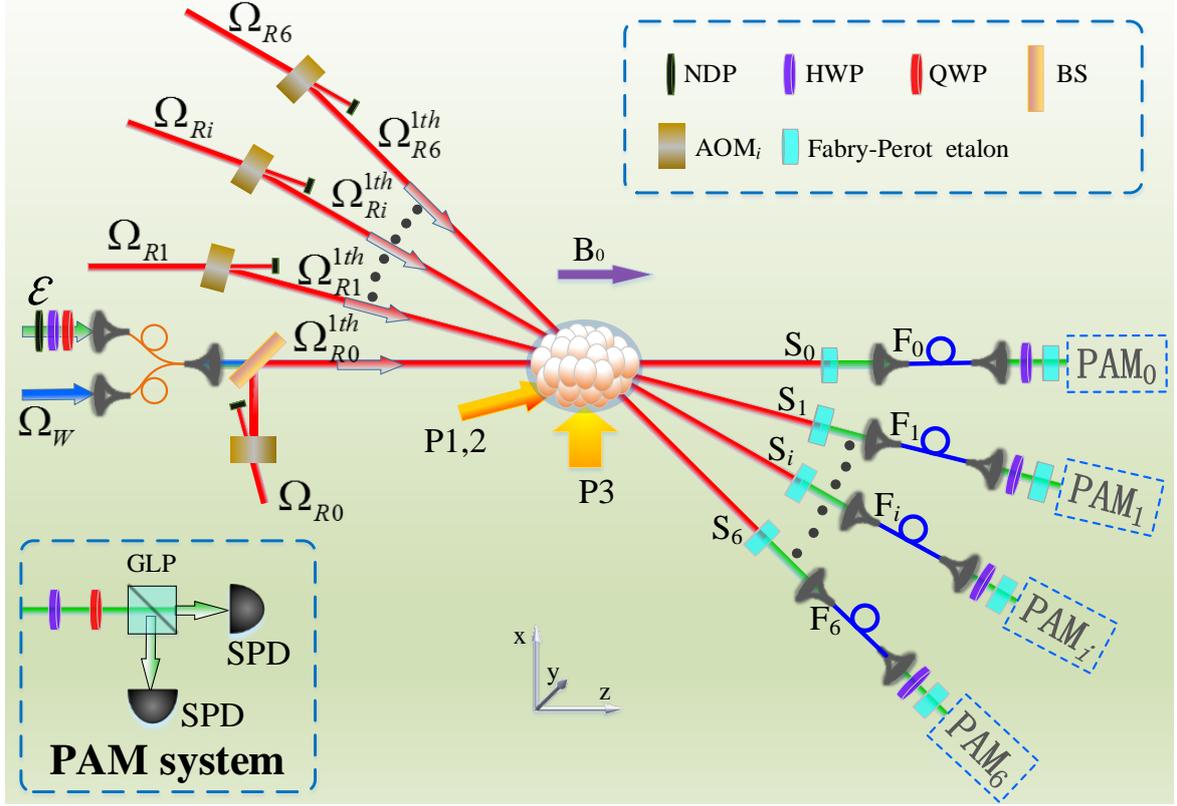

FIG. 2 (color online) Experimental setup. P1 and P2: right-circularly and left-circularly polarized pump laser beam, respectively;. P3: linearly polarized pump laser beam; σ: signal light beam; $\Omega_W$: writing light beam; $\Omega_{R0}^{1th}$, $\Omega_{R1}^{1th}$, $\Omega_{R2}^{1th}$, $\Omega_{R3}^{1th}$, $\Omega_{R4}^{1th}$, $\Omega_{R5}^{1th}$, and $\Omega_{R6}^{1th}$: 1-th order diffractions of the reading beams $\Omega_{R0}$, $\Omega_{R1}$, $\Omega_{R2}$, $\Omega_{R3}$, $\Omega_{R4}$, $\Omega_{R5}$, which propagate along the directions at the angles of $\theta = 0°$, $\theta = 0.4°$, $\theta = 0.8°$, $\theta = 2°$, $\theta = 3°$, $\theta = 4°$, and $\theta = 5°$, respectively; NDP: neutral density filters; FBS: fiber-beam-splitter; BS: polarization-insensitive beam splitter; AOM: acousto-optic modulator. F: single-mode optical fiber. GLP: the Glan-laser polarizer; PAM system: polarization analyzing and measuring (PAM) system, SPD: single-photon detector.

## 3. Experimental setup

The experimental setup is shown in Fig.2. The atomic ensemble is a cigar-shaped cloud of cold $^{87}$Rb atoms which is provided by a two-dimension (2D) magneto-optical trap (MOT). The size of the cloud of cold atoms is about 4mm×4mm×7.5mm. The input signal and writing light beams are combined with a fiber-beam-splitter (FBS), which has a 95% (5%) transmission for the signal (writing) beam. Before arriving FBS, the signal beam goes through neutral density filters, a quarter-wave plate (QWP) and a half-wave plate (HWP). The neutral density filters are used to reduce the intensity of the coherent signal light pulse. The polarization state of the signal light can be arbitrarily set by adjusting QWP and HWP. After the FBS, the signal and writing beams collinearly propagate through the cold atoms along z-direction. The pumping lasers P1, P2 and P3 are used to prepare cold atoms into the Zeeman sublevels $|a, m=\pm 1\rangle$ with equal probability, and their frequencies, polarizations and propagating directions are the same as that in the Ref. [7]. Their spot sizes (powers) in the center of cold atoms are ~7 mm (~10 mW), ~7 mm (~10 mW), 20 mm (~4mW), respectively. We use several reading beams ($\Omega_{R0}$ ... $\Omega_{Ri}$ ... $\Omega_{R6}$) which respectively propagate along different directions to retrieve the stored PPQ. The switch-on of each reading beam $\Omega_{Ri}$ is controlled by an acousto-optic modulator (AOM$_i$) and the first-order output $\Omega_{Ri}^{1th}$ of the reading beam $\Omega_{Ri}$ from AOM$_i$ is directed into the spatial mode (channel) $S_i$ with the angle $\theta_i$ relative to the z-axis. The power of each reading beam is 15mW, whose spot size is ~4 mm. When the reading beam light pulse $\Omega_{Ri}^{1th}$ illuminates the atoms, the stored SWs are converted into the retrieved signal photons, which goes into the channel $S_i$ and then is collected by a single-mode optical fiber $F_i$. Before the optical fiber $F_i$, a Fabry-Perot (FP) etalon is placed in the path to block the reading beam into the optical fiber $F_i$. After the fiber $F_i$, a HWP is used to compensate the relative phase between the retrieved $\sigma^+$-polarized and $\sigma^-$-polarized photons [7]. Passing through the HWP, the retrieved photons passes through 4 FP etalons and then are sent to a polarization analyzing and measuring (PAM$_i$) system for observing polarization fidelity of the quantum state. The total transmission of the 5 FP etalon is 58% for the signal light and ~$10^{-13}$ for the writing/reading light. The polarization analyzing and measuring system is used to measure the polarization fidelities of PPQs, which is the same as that in Ref. [7].

The experiments of storages and retrievals of the signal field are carried out in a cyclic fashion with a repetition frequency of 20Hz. In each cycle, the 87Rb atoms are trapped into the magneto-optical trap (MOT) for 42ms. After which, a bias magnetic field $B_0$ is switched on for a duration of 0.3 ms to reach 12.5 G and then the pumping lasers P1, 2, 3 and the writing laser with a power of 2.5mW are turned on. Keeping the optical pumping for 5 µs, the most of atoms have been prepared into the states $|a_{m=1}\rangle$ or $|a_{m=-1}\rangle$ with equal populations and the optical depth for the transition $|a\rangle \leftrightarrow |e\rangle$ is ~10. After the pumping, i.e., at the time $t$=0, the signal pulses (PPQs) with a pulse length of 100 ns are switched on and stored into the cloud of cold atoms by switching off the writing beam. Waiting for a storage time $t$, the stored SWs are retrieved by controllably switching on the reading beam $\Omega_{Ri}^{1th}$ ($i = 0,1...6$).

## 4. Experimental results

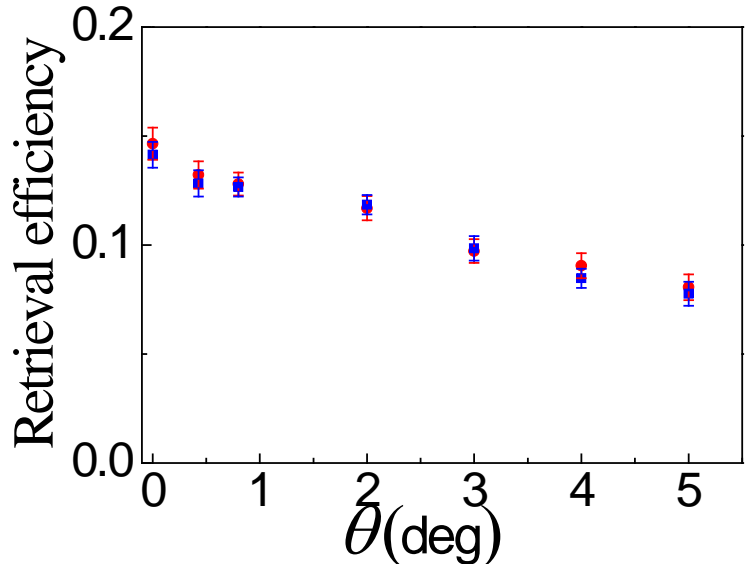

FIG. 3 (color online) Retrieval efficiencies of the signal field as the function of the angle $\theta$ at a storage time of $t = 5\mu s$, respectively. The red-circle and blue-square dots are the experimental data of the $\sigma^+$- and $\sigma^-$- polarized input signal light, respectively.

First, we measure the retrieval efficiencies of the $\sigma^\pm$-polarized components of the signal field as the function of the angle $\theta$ at a storage time of $t = 5\mu s$. The measurements are carried out when the signal-beam input peak power is 25 μW and the retrieved signal pulses are detected by photodiode detectors. The red-circle and blue-square dots in Fig.3 are the measured results of the $\sigma^+$-polarized and $\sigma^-$-polarized components of the signal field, respectively. At $\theta = 0°$, the retrieval efficiency reach to its maximum value $R_{\theta=0°}(0) = 14\%$. With increasing in the angle $\theta$, the retrieval efficiency $R_\theta$ decreases. For $\theta = 5°$, the retrieval efficiency $R_{\theta=5°}(0)$ reduces to ~8%. We attribute such reduction to spatial-mode imperfect overlap (walk-off) between the stored SWs and the single-mode fiber $F_i$, which increases as the angle $\theta$ increases.

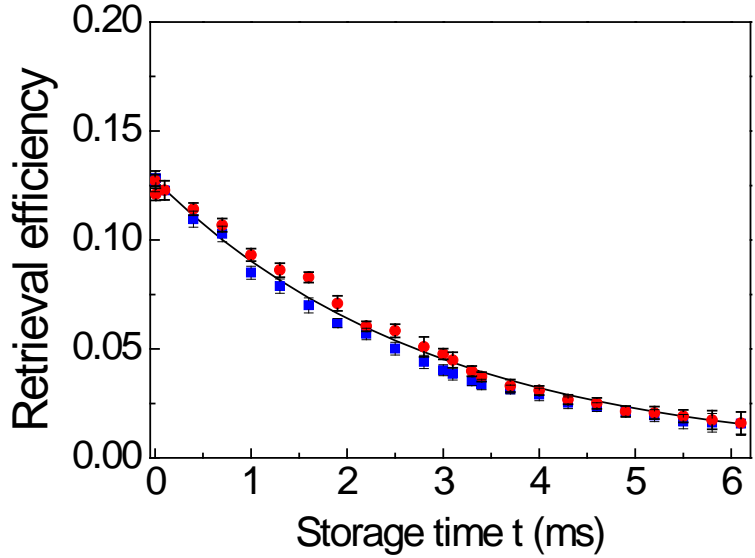

Fig. 4 (color online) Retrieval efficiencies as the function of the storage time $t$ for $\theta = 0.8°$ ($S_2$). Red circular and blue square points are the measured retrieval efficiencies of the $\sigma^+$- and $\sigma^-$-polarized input signal light, respectively. The black solid lines is the fittings to the experimental data according to $R_{\theta=0.8°}(t) = R_{\theta=0.8°}(0)\exp(-t/\tau)$ with $R_{\theta=0.8°}(0) = 12.7\%$, yielding a the storage lifetime $\tau$ =2.9ms.

Next, we measure the retrieval efficiencies of $\sigma^\pm$-polarized signal fields as the function of storage time at a fixed angle of $\theta = 0.8°$. The red circle and blue square dots in Fig.4 are the measured results, respectively, and the black solid line is the fittings to the experimental data according to $R_\theta(t) = R_{\theta=0.8°}(0)e^{-t/\tau}$ with $R_{\theta=0.8°}(0) = 12.7\%$, which yield a storage lifetime of ~2.9ms. Such lifetime is longer than that (1.5ms) in the previous work [7]. The main reason is that the magnetic-field gradient of ~35mG in the previous work [7] is reduced to 5mG/cm in the presented work.

Subsequently, we measure the polarization fidelities of the retrieved PPQ for several different angles $\theta$. The input signal pulse is decreased to the single-photon level (i.e., the mean photon $\bar{n}=1$) by using the neutral density filters. $\bar{n}$ is determined by detecting the probability per pulse at the case without the cold atomic cloud and considering the total detection efficiency of $\eta_d = 23\%$, which includes the coupling efficiency of the single-mode fiber (80%), total transmission for the 5 Fabry-Perot etalons (58%), the efficiency of multi-mode fiber coupling to single-photon detectors SPD1 or SPD2 (97%) and the quantum efficiency of the single-photon detectors SPD1 and SPD2 (50%).

The quality of storage and retrieval of PPQ can be characterized by quantum process matrix $\chi$, which can be obtained according to the relation [29, 30]:

$$\rho_{out} = \sum_{m,n=0}^{3} \chi_{mn} \sigma_m \rho_{in} \sigma_n^{\dagger} \qquad (4)$$

where $\sigma_i$ are Pauli operators, $\rho_{in}$ and $\rho_{out}$ are the density matrixes of the input signal photons and retrieved signal photons, respectively. For obtaining the density matrix $\rho_{out}$, we follow three steps: performing the experiments of the storage and retrieval for four input polarization states $|H\rangle$, $|V\rangle$, $|D\rangle$ and $|R\rangle$, respectively, analyzing the retrieval photon in three mutually-unbiased bases $|H\rangle-|V\rangle$, $|R\rangle-|L\rangle$ and $|D\rangle-|A\rangle$, reconstructing the density matrix $\rho_{out}$ by means of the quantum state tomography [31], where *H, V, R, L, D* and *A* denote horizontal, vertical, right circular, left circular, diagonal (45°), and antidiagonal (-45°) polarizations, respectively. Based on the obtained density matrix $\rho_{out}$, we reconstruct the matrix $\chi$ and then obtain quantum-process fidelity $F_{Process}$ according to the definition $F_{process} = Tr\left(\sqrt{\sqrt{\chi}\chi^{ideal}\sqrt{\chi}}\right)^2$ with $\chi_{0,0}^{ideal} = 1$ [29]. At a storage time of $t = 5\mu s$, we measure the quantum-process fidelities $F_{Process}$ for several different angles $\theta$, the results are listed in Table 1, which show that they are not less than 89%. At $\theta = 0.8°$, we also measure the quantum process fidelity $F_{Process}$ as the functions of storage time *t*. The square dots in Fig. 5 are shown the measured results, which decreases with storage time. We attribute the decrease to the following two factors. First, since the retrieval efficiency exponentially reduces with the storage time, the background noise gradually becomes a main contribution to the single-photon-counting events which make the polarization fidelity reduce. Second, the dephasing between the two spin waves $\psi^{\pm}$, induced by the temporal fluctuations of the magnetic field in the z-direction will decrease the polarization fidelity also. Taking into account the two factors, the $F_{Process}$ can be approximately expressed as [7]: $F_{Process} \approx \frac{(1+\gamma(t))\eta_d R_\theta(t) + N}{2(\eta_d R_\theta(t) + 2N)}$, where $\eta_d = 23\%$ is the total efficiency for detecting the retrieved signal photons, $N = 7\times10^{-4} / pulse$ corresponds to the background-noise photon number per pulse, $R_\theta(t) = R_{\theta=0.8°}(0)\exp(-t/\tau)$ is the retrieval efficiency for $\theta = 0.8°$, $\gamma(t) = \exp[-t^2/\sigma_\gamma^2]$ is the dephasing factor, which results from the dephasing between the two spin waves $\psi^\pm$. The solid curve in Fig.5 is the fitting to the data of $F_{Process}$, which yields the $e^{-1}$ dephasing time of $\sigma_\gamma = 104\ ms$, corresponding to a magnetic-field temporal fluctuations of $\sigma_B \approx 0.4\ mG$ [7]. From the Fig.5, one can see that the measured quantum-process fidelity reach its maximal value of 94.2% at storage time $t= 0.85$ms and is still beyond 78% at storage time $t=6$ ms.

**Table 1** Quantum process fidelities $F_{process}$ for different angles $\theta$ at a storage time of $t = 5\mu s$. The errors have been obtained by Monte Carlo simulation taking into account the statically uncertainty of photon counts.

| Angle | $\theta = 0°$ ($S_0$ mode) | $\theta = 0.4°$ ($S_1$ mode) | $\theta = 0.8°$ ($S_2$ mode) | $\theta = 2°$ ($S_3$ mode) | $\theta = 3°$ ($S_4$ mode) | $\theta = 4°$ ($S_5$ mode) | $\theta = 5°$ ($S_6$ mode) |
|---|---|---|---|---|---|---|---|
| $F_{process}$ (%) | 90.2±2.6 | 90.3±1 | 91.4±1.4 | 90.6±2.3 | 91±2 | 89.1±1.8 | 89.5±2.4 |

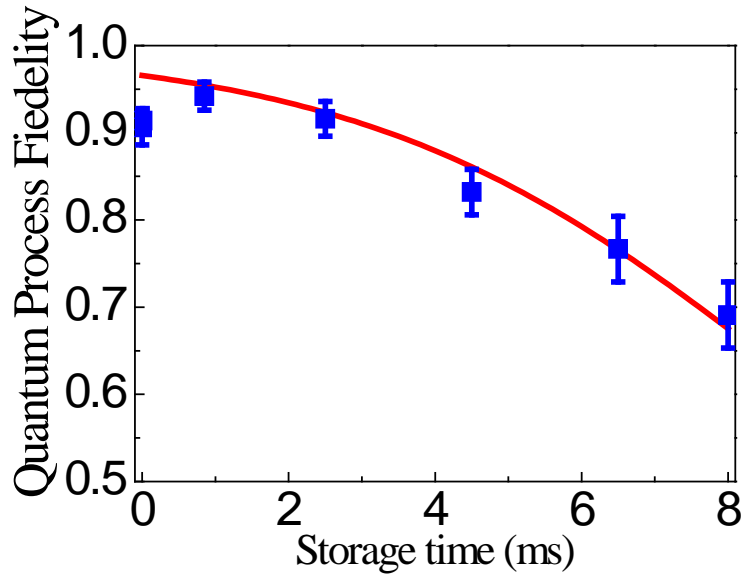

Fig. 5 (color online) Quantum process fidelity as the function of the storage time $t$ at $\theta = 0.8°$. The data are fitted by the expression $F_{process} = \frac{(1+\gamma(t))\eta_d R_\theta(t) + N}{2(\eta_d R_\theta(t) + 2N)}$ with the measured total detection efficiency $\eta_d \approx 0.23$ and background noise $N = 7 \times 10^{-4} / pulse$. The retrieval efficiency $R_\theta(t) = R_{\theta=0.8°}(0)\exp(-t/\tau)$ with $R_{\theta=0.8°}(0) = 12.7\%$. The fitting yields the $e^{-1}$ dephasing time of $\sigma_\gamma = 104 ms$. The errors have been obtained by Monte Carlo simulation taking into account the statically uncertainty of photon counts.

## 5. Conclusion

In summary, we have demonstrated an experiment in which the long-lived quantum memories for photonic polarization qubits (PPQs) can be controllably released into 7 different photonic output channels respectively. The measured quantum process fidelity of the retrieved PPQ for each of the channels is not less than 89%. The measured maximum quantum-process fidelity for the retrieved single-photon polarization states is 94.2% for storage time of $t = 0.85ms$. For storage time of 6 ms, the measured process fidelity of the retrieved PPQs is still higher than the threshold of 78% for the violation of the Bell inequality [32]. The retrieval efficiency for a zero storage time is 14%, which can be further improved by either increasing the optical depth of the cold atoms [8, 33] or coupling the atoms into an optical cavity [6, 34]. Base on the mechanism of the controllable release of stored PPQs, one can build quantum memory elements capable of routing retrieved photon qubits and then can find applications in quantum information processing based on quantum internet.

**Acknowledgments**

We acknowledge funding support from the 973 Program (2010CB923103), the National Natural Science Foundation of China (No.10874106, 11274211, 60821004), the Program for Sanjin Scholars of Shanxi Province of China.